\documentclass[twocolumn,showpacs,nofootinbib,preprintnumbers,amsmath,amssymb]{revtex4}

\usepackage[utf8]{inputenc}
\usepackage{bm}
\usepackage{color}
\usepackage{hyperref}
\usepackage{amsmath, mathtools}
\usepackage{amssymb}
\usepackage{wasysym}
\usepackage{graphicx} 
\usepackage{booktabs} 
\usepackage{pdfsync}
\usepackage{verbatim}
\usepackage{latexsym}
\usepackage{dsfont}

\def\sec#1{\section{#1} }
\def\ssec#1{\subsection{#1} }
\def\sssec#1{\subsubsection{#1} }

\def\({\left(}
\def\){\right)}
\def\[{\left[}
\def\]{\right]}

\def\a{\alpha}

\def\f#1#2{\frac{#1}{#2}}
\def\g{\gamma}

\def\d{\partial}
\def\de{\delta}

\def\ep{\epsilon}

\def\k{\kappa}

\def\n{\nu}
\def\o{\omega}

\def\p{\pi}

\def\t{\tau}
\def\th{\theta}

\def\<{\langle}
\def\>{\rangle}

\providecommand{\abs}[1]{\left\lvert#1\right\rvert}

\usepackage{graphicx}
\usepackage{dcolumn}
\usepackage{bm}

\begin{document}

\preprint{APS/123-QED}

\title{A perturbative method for resolving contact interactions in quantum mechanics}

\author{David M. Jacobs}
\affiliation{%
Physics Department, Hamilton College\\
198 College Hill Rd., Clinton, NY 13323
}%

\date{\today}

\begin{abstract}
Long-range effective methods are ubiquitous in physics and in quantum theory, in particular. Furthermore, the reliability of such methods is higher when the nature of short-ranged interactions need not be modeled explicitly. This may be necessary for two reasons: (1) there are interactions that occur over a short range that cannot be accurately modeled with a potential function and/or (2) the entire Hamiltonian loses its reliability when applied at short distances. This work is an investigation of the utility and consequences of omitting a finite region of space from quantum mechanical analysis, accomplished by imposition of an artificial boundary behind which obscured short-ranged physical effects may operate. With this method, a free function of integration that depends on momentum is interpreted as a function encoding information needed to match a long-distance wavefunction to an appropriate state function on the other side of the boundary. Omitting part of the space from analysis implies that the strict unitarity requirement of quantum mechanics must be relaxed, since particles can actually propagate beyond the boundary. Strict orthogonality of eigenmodes and hermiticity of the Hamiltonian must also be relaxed in this method; however, all of these canonical relations are obeyed when averaged over sufficiently long times.  What is achieved, therefore, appears to be an effective long-wavelength theory, at least for stationary systems. As examples, the quantum defect theory of the one-dimensional Coulomb interaction is recovered, as well as a new perspective of the inverse-square potential and the free particle, as well as the Wigner time delay associated with contact interactions. Potential applications of this method may include three-dimensional atomic systems and two-dimensional systems, such as graphene. 

\end{abstract}

\pacs{Valid PACS appear here}
\maketitle

\sec{Introduction}\label{Sec:intro}

Several methods are known to effectively describe short-distance physical effects in quantum mechanics. The Dirac delta function potential is the most well-known example, but it has limitations. It is useful only to the extent that the range of the potential can be approximated to be zero, and it is usually used in the context of a perturbative technique in which unperturbed wavefunctions are used to compute its effects.  The method of self-adjoint extensions is arguably an improvement upon this; it works where the delta-function technique fails or, at least, requires a complicated infinite renormalization and doesn't rely on the standard perturbative framework \cite{jackiw1995diverse}. In that method a non-trivial boundary condition can model a contact interaction, see e.g. \cite{Bonneau:1999zq, PhysRevA.66.052102, Roy:2009vc,  BeckThesis, Thompson:2018dge}.

The Dirac delta and self-adjoint extension methods still fail in particular cases, such as free particles obeying the Dirac equation in two and three dimensions, for example. Furthermore, even in the cases where the two techniques work and agree in their predictions, they are only capable of describing leading order effects. The method of self-adjoint extensions is also limited because, in many systems, the need to normalize the eigenstates results in a trivial boundary condition; for example, $\psi(0)=0$ for any $\ell\neq0$ solution of the Schrodinger-Coulomb equation, hence only $s$-waves can have non-trivial interactions.

There are techniques for capturing short-ranged effects for all $\ell$ channels; see for example, reference \cite{Lepage:1997cs} in which ultraviolet (UV) corrections to the three-dimensional Schrodinger-Coulomb system are dealt with in a perturbative fashion. In this approach, one explicitly models the UV effects with, essentially, a series of momentum-dependent contact potentials. However, one might question the general validity of such a method, for example, down to distances where the non-relativistic Schrodinger equation should lose its predicative power, i.e. where $\abs{V(r)}\gtrsim mc^2$. 

We therefore pose the question: can a reliable long-distance effective description be constructed that does not rely on an explicit model of how the Hamiltonian deviates from its long-distance form? \emph{Can one completely omit from analysis the region over which short-ranged interactions operate?}

A first attempt toward this goal was presented in \cite{Jacobs:2015han}. In that work a small region of space, bounded artificially, was excluded from analysis. Since observables cannot depend on what volume of space is excluded, the wavefunction boundary conditions run, in the renormalization group (RG) sense, with the boundary radius. However, in that work, the boundary radius had to be taken to zero to ensure that the Hamiltonian was hermitian and that unitarity is strictly obeyed. Burgess \emph{et al.} followed a similar path, using effective field theory arguments to derive the form that the boundary conditions must take at the origin \cite{Burgess:2016lal, Burgess:2016ddi}. They have considered the $1/r^2$ and Coulomb potentials in three dimensions, focusing on an effective description of $s$-states.

The present article is an extension to \cite{Jacobs:2015han}, and is about the utility and consequences of \emph{not} taking the boundary radius to zero. What results is a long-wavelength effective theory, applicable to a class of stationary quantum systems, that captures short-ranged effects perturbatively.   It is distinct from other methods, e.g. such as that of reference \cite{Lepage:1997cs}, in that the Hamiltonian is not be specified at short distances. Its robustness is also demonstrated in the recovery of known results for systems in which short-distance phenomena result in long-distance effects, e.g. quantum defect theory that describes Rydberg atoms \cite{Seaton_1983}.  

For illustration of the method proposed, we will limit discussion to those systems in which a particle propagates on an infinite half-line with coordinate $x\in [0,\infty)$ and whose evolution is dictated by a Hamiltonian of the non-relativistic form
%
\begin{equation}\label{generic_Ham}
H=-\f{\hbar^2}{2m}\f{\d^2}{\d x^2} + V(x)\,,
\end{equation}
where $m$ will be reserved for the particle mass, or the reduced mass of a two-particle system. The point $x=0$ represents a hard physical boundary that may correspond to the edge of the system in which a particle is contained, the point of contact between two particles, or to the origin of coordinates in a higher-dimensional system.
In order to capture unknown short-distance effects in the vicinity of $x=0$, an artificial boundary is placed at $x=x_\text{b}$ so that the region $0\leq x<x_\text{b}$ is no longer in the domain of analysis. Although the strict unitarity requirement will be relaxed in this work, the norm of each eigenfunction, $\psi_i$ is conserved by ensuring that its associated probability current density vanish at the boundary, i.e.
\begin{equation}\label{boundary_current}
\psi_i'^\star (x_b)\psi_i(x_b) - \psi_i^\star (x_b)\psi_i'(x_b)=0\,.
\end{equation}
Following reference \cite{Bonneau:1999zq}, we may use the identity
\begin{equation}
\(x\bar{y}-\bar{x}y\)=\f{i}{2}\(\abs{x+iy}^2-\abs{x-iy}^2\)
\end{equation}
to write the condition \eqref{boundary_current} as
\begin{equation}\label{new_condition}
\abs{\psi_i(x_\text{b})+iw \psi_i'(x_\text{b})}^2 - \abs{\psi_i(x_\text{b})-iw \psi_i'(x_\text{b})}^2=0\,,
\end{equation}
where $w$ is an arbitrary real-valued constant with units of length and is only inserted for dimensional reasons. The two terms whose absolute values are taken in \eqref{new_condition} are apparently equal up to a phase factor; it follows that the general boundary condition is therefore
\begin{equation}\label{1d_standard_bc}
\psi_i(x_\text{b}) + Z(x_b)\, \psi_i'(x_\text{b})=0\,,\notag
\end{equation}
where the boundary function $Z(x_b)$ can take any real value (see also reference \cite{Jacobs:2015han}).

What is new in this work is to promote the boundary function to be unique to the eigenmode, that is, $Z\to Z_i$ so that
\begin{equation}\label{1d_EQM_bc}
\psi_i(x_\text{b}) + Z_i(x_b)\, \psi_i'(x_\text{b})=0\,.
\end{equation}
Equation \eqref{1d_EQM_bc} is the central equation to this work. By demanding that observables do not depend on $x_b$, a differential (RG) equation can be derived whose solution contains an integration function, constant with respect to $x_b$ but with arbitrary dependence on momentum. A simple perturbative \emph{ansatz} for this function, here called $\chi$, is remarkably effective at modeling a systems's long-distance behavior. In the very low-energy limit, as momentum approaches zero, the results coincide with that of the method of self-adjoint extensions, such as in references  \cite{PhysRevA.66.052102,  BeckThesis, Jacobs:2015han, EssinGriffiths2006}.

In Sections \ref{Sec:1OVERx}, \ref{Sec:1OVERx2}, and \ref{sec:free} the one-dimensional Coulomb, $1/x^2$, and free particle systems are considered, respectively. Bound state eigenvalues and scattering phase shifts are computed with the proposed effective method and compared to a specific UV-complete model in which the potential near to the origin is constant. In Section \ref{Sec:WignerDelay} the Wigner time delays are computed for these systems within the context of this method. In Section \ref{Sec:Consequences} the issues of orthogonality, hermiticity, and unitarity are addressed and it is shown how these canonical relations are recovered after averaging over sufficiently long times. We conclude in Section \ref{Sec:Discussion} with a summary and discussion of possible applications.

\sec{The $1/x$ potential}\label{Sec:1OVERx}
Consider a particle on the half-line subject to evolution dictated, at long distances, by the Hamiltonian
\begin{equation}\label{1dCoulombHam}
H=-\f{\hbar^2}{2m}\f{\d^2}{\d x^2}+\f{\a \hbar c q_1 q_2}{x}\,,
\end{equation}
where $\a=e^2/(4\pi \ep_0 \hbar c)$ and $q_1$ and $q_2$ are the charges of two objects involved; we refer to this as the one-dimensional Coulomb system. Setting $\hbar=c=1$ and defining
\begin{equation}
\k\equiv m\a q_1 q_2\,,
\end{equation}
let
\begin{equation}
2mE\equiv
\begin{cases}
-q^2, &(E<0)\\
k^2, &(E>0)\,.
\end{cases}
\end{equation}

For bound states ($E<0$), let the solutions to the Schrodinger eigenvalue problem be
\begin{equation}
\psi(x)=e^{-q x} x g(x)\,.
\end{equation}
The Schrodinger equation with \eqref{1dCoulombHam} as the Hamiltonian then yields
\begin{equation}
xg''(x) + 2 \(1-qx\)g''(x) - 2\(q+\k\)g(x)=0\,.
\end{equation}
One set of linearly independent\footnote{    This set is linearly independent so long as $q/\k$ is not equal to a negative integer, an explicit assumption that we make. In a real system there is zero probability that this would occur. In any case, the other linearly-independent solution that can be found also cannot be normalized, making the point moot.}      solutions to this equation are the confluent hypergeometric functions $U(1+\f{\k}{q} | 2 | 2qx) $ and $_1F_1(1+\f{\k}{q} | 2 | 2qx)$. Normalizeability will require that the 2nd solution be omitted, therefore
\begin{equation}\label{oneOVERx_sol_bound}
\psi=A e^{-qx} x\, U\(1+\f{\k}{q} \Big| 2 \Big| 2qx\)  \,,
\end{equation}
where $A$ is a normalization factor.  The spectrum of $q$ are observable.

For scattering states ($E>0$) one set of solutions is $e^{-ik x} x$ times a linear combination of $U(1-\f{i\k}{k} \big| 2 \big| 2ikx) $ and $_1F_1(1-\f{i\k}{k} \big| 2 \big| 2ikx)$. The choice
\begin{align}\label{oneOVERx_sol_scatt}
\psi=A e^{-ikx} x\, \( \psi_L - e^{2i\de}\psi_R   \) \,,
\end{align}
where
\begin{align}
\psi_R= e^{-\frac{\pi  \kappa }{2 k}} \Gamma \left(1-\frac{i \kappa }{k}\right) \Bigg(\,
   _1F_1\left(1-\frac{i \kappa }{k} \Big| 2 \Big| 2 i k x\right)\notag\\
   \left.+\frac{e^{\frac{\pi  \kappa }{k}}
   U\left(1-\frac{i \kappa }{k} \Big| 2 \Big| 2 i k x\right)}{\Gamma \left(\frac{i \kappa
   }{k}+1\right)}\right)\,,
\end{align}
and
\begin{equation}
\psi_L=e^{\frac{\pi  \kappa }{2 k}} U\left(1-\frac{i \kappa }{k} \Big| 2 \Big| 2 i k x\right)\,,
\end{equation}
gives the asymptotic form
\begin{equation}
\lim_{x\to\infty}\psi \sim A\, \( e^{-ikx + i\f{\k}{k}\ln{2kx}} - e^{2i\de}e^{+ikx - i\f{\k}{k}\ln{2kx}}   \) \,,
\end{equation}
where $2\de$ is the total phase shift, at a particular value of $x$, for an incoming wave ($\psi_L$) scattered toward positive $x$ ($\psi_R$).

\ssec{Effective Model}
\sssec{Bound state ($E<0$) solutions}
 Application of the boundary condition, equation \eqref{1d_standard_bc}, and expanding it to lowest order in $qx_b$ can be written
\begin{equation}\label{Coulomb_Digamma_1}
\Psi\(1+\f{\k}{q}\)=-\f{1}{2\k Z(x_b)}-\ln{2qx_b}+\f{q}{2\k}-2\g\,,
\end{equation}
where $\g=0.577\dots$ is the Euler–Mascheroni constant and  $\Psi(x)$ refers to the digamma function,
\begin{equation*}
\Psi(x)\equiv \f{\Gamma'(x)}{\Gamma(x)}\,.
\end{equation*}
 As the left-hand side of \eqref{Coulomb_Digamma_1} must be independent of $x_b$, it follows that the boundary function has the form
\begin{equation}
Z(x_b)=\(\chi(q^2)-2\k\ln{\f{x_b}{b_0}}\)^{-1}\,,
\end{equation}
where $\chi(q^2)$ is an arbitrary function of $q^2$, and the parameter $b_0$ is an arbitrary constant, independent of $q$. That $\chi$ is a function of $q^2$ is dictated by the form of the Schrodinger equation which must be valid for some finite distance behind the artificial boundary.  It follows that
\begin{equation}\label{1dCoulomb_EQM_quant_1}
\Psi\(1+\f{\k}{q}\)=-\f{\chi(q^2)}{2\k}-\ln{2qb_0}+\f{q}{2\k}-2\g\,.
\end{equation}

Motivated by the known spectrum in the 3-dimensional case, we make the bound state \emph{ansatz}
\begin{equation}
q=-\f{\k}{n-\de}\,,
\end{equation}
where $n$ is an integer and in this context $\de$ is called the \emph{quantum defect} (see, e.g., \cite{hartree_1928, Seaton_1983}) . In general, there is no reason to expect that $\de$ should be small, a fact that would be useful for a perturbative analysis; however we can define 
\begin{equation}
n-\de\equiv \tilde{n}-\tilde{\de}\,,
\end{equation}
where $\tilde{n}$ is the integer closest to $n-\de$, and $\tilde{\de}$ is the remaining fractional part, obeying $\abs{\tilde{\de}}<1/2$ by definition. With the simplifying choice
\begin{equation}\label{Coulomb_b0_choice}
b_0=-\f{1}{2\k}e^{-2\g}\,, 
\end{equation}
it follows from equation \eqref{1dCoulomb_EQM_quant_1} that
\begin{equation}\label{1dCoulomb_quant_2}
\Psi\(1-\tilde{n}+\tilde{\de}\)=-\f{\chi(q^2)}{2\k}+\ln{\(\tilde{n}-\tilde{\de}\)}-\f{1}{2\(\tilde{n}-\tilde{\de}\)}\,.
\end{equation}
Using the reflection formula, the digamma function may be written
\begin{align}
\Psi\(1-\tilde{n}+\tilde{\de}\)= -\p \cot{\p\tilde{\de}}  + \Psi\(\tilde{n}-\tilde{\de}\)\,.\notag
\end{align}
Making the notational choice
\begin{equation}\label{nu_def}
\n_{\tilde{n}}\equiv \tilde{n}-\tilde{\de}\,,
\end{equation}
we expand in small $\tilde{\de}$ and large $\n_{\tilde{n}}$, for which
\begin{equation}
-\p \cot{\p\tilde{\de}}\sim -\f{1}{\tilde{\de}} + \f{\p^2}{3}\tilde{\de} + {\cal O}\(\tilde{\de}^3\)\notag\,,
\end{equation}
and
\begin{equation}
\Psi\(\n_{\tilde{n}}\) \sim \ln{\n_{\tilde{n}}}-\f{1}{2\n_{\tilde{n}}} -\f{1}{12\n_{\tilde{n}}^2}  + {\cal O}\(\n_{\tilde{n}}^{-4}\)\,.\notag 
\end{equation}
It then follows from equation \eqref{1dCoulomb_quant_2} that
\begin{equation}
\tilde{\de}^{-1} =\f{\chi(q^2)}{2\k}
 + \f{\p^2}{3}\tilde{\de}     -\f{1}{12\n_{\tilde{n}}^2}  + {\cal O}\(\n_{\tilde{n}}^{-4}\) + {\cal O}\(\tilde{\de}^3\)\,.
\end{equation}

We have up to this point said nothing about the form of $\chi(q^2)$. However, if there is data that indicates $\tilde{\de}$ approaches a constant for very large $\tilde{n}$, as is the case for real three-dimensional atoms, $\chi(q^2)$ should obey
\begin{equation}
\lim_{q\to0}\chi(q^2)=c_0\,,
\end{equation}
for some momentum scale $c_0$. If deviations can be described analytically, at least for large $\tilde{n}$, we expect there to be an approximant that can be written in terms of $q^2$, as described above. It appears simplest to posit the series form
\begin{equation}\label{Coulomb_effective_ansatz}
\chi(q^2)=c_0 + c_2 q^2  + {\cal O}(q)^4\,,
\end{equation}
from which it follows
\begin{align}
\tilde{\de}^{-1} =\f{c_0}{2\k}&\(1+\f{2\k}{c_0}\(\f{c_2 \k}{2}-\f{1}{12}\)\n_{\tilde{n}}^{-2}+\f{2\p^2\k}{3c_0}\tilde{\de} \) \notag\\
&+ {\cal O}\(\n_{\tilde{n}}^{-4}\) + {\cal O}\(\tilde{\de}^3\) \notag\,.
\end{align}
This may be perturbatively solved for $\tilde{\de}$ and written in the more familiar form
\begin{equation}\label{Schr-Coul_tildede_result}
\tilde{\de}\simeq\tilde{\de}_0+\f{\tilde{\de}_2}{\n_{\tilde{n}}^2} + {\cal O}\(\n_{\tilde{n}}^{-4}\)\,,
\end{equation}
where
\begin{align}\label{Schr-Coul_tildede_result_2}
\tilde{\de}_0&=\f{2\k}{c_0}\(1+\f{\p^2}{3}\(\f{2\k}{c_0}\)^2\)^{-1}\notag\\
\tilde{\de}_2&=-\(\f{2\k}{c_0}\)^2\[\f{c_2\k}{2} - \f{1}{12} \]\(1+\f{\p^2}{3}\(\f{2\k}{c_0}\)^2\)^{-1}\,.
\end{align}

In summary, the observable energy eigenvalues labelled by integer $\tilde{n}$ are given by
\begin{equation}
E_{\tilde{n}}=-\f{\k^2}{2m}\f{1}{\n_{\tilde{n}}^2}
\end{equation}
where $\n_{\tilde{n}}$ is given by equations \eqref{nu_def}, \eqref{Schr-Coul_tildede_result} and \eqref{Schr-Coul_tildede_result_2}. Workers that study Rydberg atoms will recognize this result as equivalent to the extended Ritz formula \cite{hartree_1928, Seaton_1983}. This result confirms the power of the method proposed in this article. \emph{No model for the deviation from a pure Coulomb potential was imposed in the region behind the artificial boundary}; only a plausible series form for the free function $\chi(q^2)$ was posited. 

\sssec{Scattering ($E>0$) solutions}
With the choice of $b_0$ given in \eqref{Coulomb_b0_choice}, here we find
\begin{equation}\label{Coulomb_EQM_scatt}
e^{2i\de}=\frac{\Gamma \left(1 + \frac{i \kappa }{k}\right)}{\Gamma \left(1-\frac{i \kappa
   }{k}\right)}\(1+ f(k)\)^{-1}\,,
\end{equation}
where
\begin{equation}
f(k)=\frac{2 \pi  \kappa  \left(\coth \left(\frac{\pi  \kappa }{k}\right)-1\right)}{\pi 
   \kappa -2 i \kappa  \ln \left(-\frac{k}{\kappa }\right)-2 i \kappa  \Psi\left(-\frac{i \kappa }{k}\right)-i \chi (-k^2)+k}\,.
\end{equation}
Under the assumption that the function $\chi(q^2)$ continues analytically through zero to $q^2=-k^2$, the series form is apparently
\begin{equation}
\chi(-k^2)=c_0 - c_2 k^2  + {\cal O}(k)^4\,.
\end{equation}

\ssec{A UV-complete model}
Consider a model in which the Coulomb singularity is regulated with a potential step, parameterized as
\begin{equation}
V(x)= 
\begin{cases}
\f{\k}{m L},~~~~~&(0\leq x \leq L)\\
\f{\k}{m x},~~~~~~~&(x>L)\,.
\end{cases}
\end{equation}
We will focus on systems in which the step width is much smaller than the Coulomb length scale, i.e. $L\ll \abs{\k^{-1}}$. 

For bound states ($E<0$), define 
\begin{align}
q^2&\equiv -2mE\notag\\
p^2&\equiv-\f{2\k}{L} -q^2\,,
\end{align}
so that for $x> L$ the solutions are just as in equation \eqref{oneOVERx_sol_bound}
\begin{equation}
\psi_\text{out}=A e^{-qx} x\, U\(1+\f{\k}{q} \Big| 2 \Big| 2qx\)  \,,
\end{equation}
and within $x\leq L$
\begin{equation}\label{Coulomb_UV-complete_innersol}
\psi_\text{in}=B \sin{px}\,.
\end{equation}
Matching the wave function and its derivative at $x=L$ can be described with a single matching equation
\begin{equation}\label{UV-complete_Sch-Coul}
\psi_\text{out}'\(L\)\psi_\text{in}\(L\)-\psi_\text{out}\(L\)\psi_\text{in}'\(L\)=0\,,
\end{equation}
which may be solved numerically to find the exact energy eigenvalues of this UV-complete system. However, analytical progress is made by expanding equation \eqref{UV-complete_Sch-Coul} in both small $qL$ and $\k L$, and using the digamma recurrence relation
\begin{equation*}
\Psi(x+1)=\Psi(x)+\f{1}{x}\,.
\end{equation*}
Putting the result into the same form as equation \eqref{1dCoulomb_EQM_quant_1} yields
\begin{widetext}
\begin{align}\label{1dCoulomb_UV-complete_quant}
\Psi\(1+\f{\k}{q}\)=&\f{2053}{700}-2\gamma +\f{3}{2\(\k L\)^2} -\f{12}{5\k L} +\f{2276\k L}{7875} + {\cal O}\(\k L\)^2 -\ln{2qL}+\f{q}{2\k}\notag\\
&~~~~~~~ + \f{q^2}{31500\k^2}\(-4725+1920\k L + {\cal O}\(\k L\)^2 \) + {\cal O}\(q\)^3\,.
\end{align}
\end{widetext}
For scattering states ($E>0$), $\psi_\text{out}$ is the same as given in equation \eqref{oneOVERx_sol_scatt}, while $\psi_\text{in}$ is given in equation \eqref{Coulomb_UV-complete_innersol} with
\begin{equation}
p^2\equiv-\f{2\k}{L} +k^2\,.
\end{equation}
As in equation \eqref{UV-complete_Sch-Coul}, matching the wavefunction and its derivative, one may solve for the scattering phase shift, $e^{2i\de}$. The results are summarized in the section below.

\ssec{Matching the UV-complete \& effective models}

By matching the bound state results --  equation \eqref {1dCoulomb_UV-complete_quant} to the effective result, equation \eqref{1dCoulomb_EQM_quant_1} and choice of $b_0$ \eqref{Coulomb_b0_choice} -- the effective parameters up to ${\cal O}(\k L)$ are apparently
\begin{align}\label{Coulomb_effective_parameters}
c_0&=\k\(-\f{2053}{350}+4\g - \f{3}{\(\k L\)^2} + \f{24}{5 \k L} - \f{4552\k L}{7875} \right.\notag\\
&~~~~~~~~~~~~~+ 2\ln{\(-2\k L\)}\Bigg)\notag\\
c_2&=\f{315-128\k L}{1050\k}\,.
\end{align}

\begin{figure}[htp]
  \begin{center}
    \includegraphics[scale=.62]{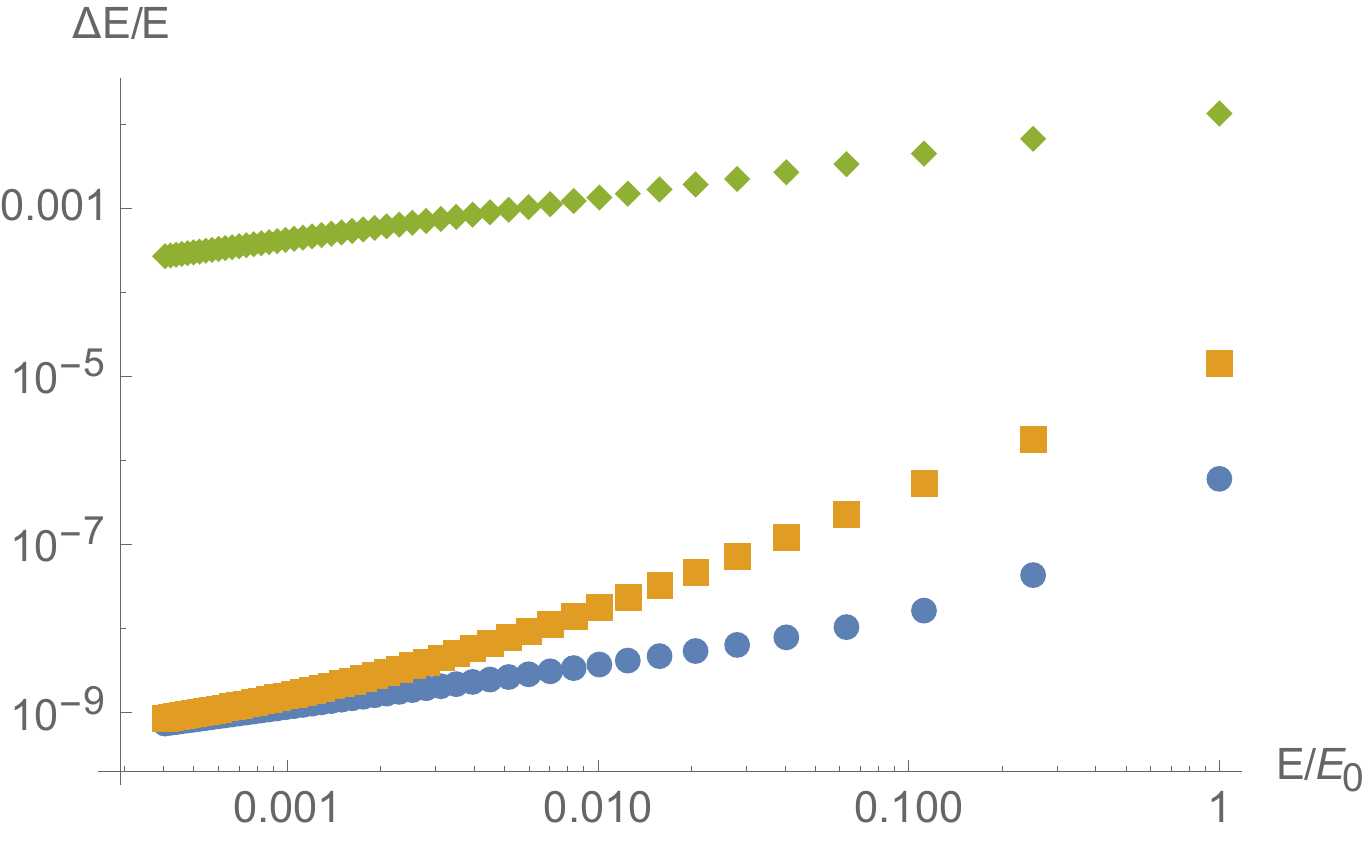}
  \end{center}
  \caption{Relative errors in the binding energies, compared to the UV-complete model, as a function of normalized binding energy. Shown are results for canonical binding energies (diamonds), the lowest order results (squares) and the effective method (circles). The parameter choice $L=0.11 \abs{\k}^{-1}$ has been made for illustration. }
  \label{RelativeEnergyDiff}
  \end{figure}

For bound states, equation \eqref{UV-complete_Sch-Coul} is solved numerically for $q$ in the UV-complete model and compared with the  effective model calculation using equations \eqref{Schr-Coul_tildede_result} and \eqref{Schr-Coul_tildede_result_2} for selected model parameters; the results are summarized in Figure \ref{RelativeEnergyDiff}, which shows the relative error in the binding energies computed in various models, compared with the actual binding energy computed in the UV-complete model. The energies on the horizontal axis are normalized to the ground state energy, $E_0$. The canonical binding energies are determined with the canonical boundary condition $\psi(0)=0$, corresponding to $c_0\to \infty$. The lowest order results are equivalent to the self-adjoint extension analysis in which $c_0$ is given in \eqref{Coulomb_effective_parameters} but $c_2=0$, whereas the effective method proposed here uses both $c_0$ and $c_2$ as given in \eqref{Coulomb_effective_parameters}.

\begin{figure}[htp]
  \begin{center}
    \includegraphics[scale=.7]{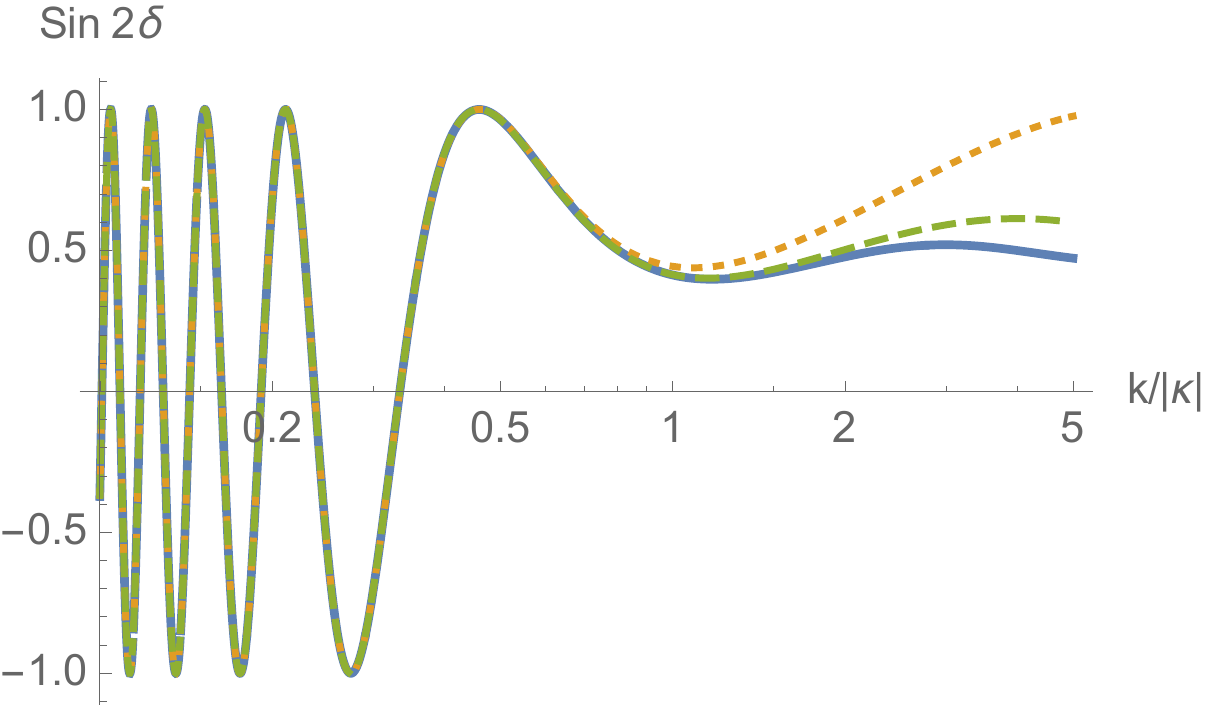}
  \end{center}
  \caption{Scattering results for the one-dimensional Coulomb system for $\k<0$.  As a function of $k$, $\sin{2\de}$ is shown for the parameter $L=0.9 \abs{\k}^{-1}$. as computed in the UV-complete model (solid curve), the lowest order (LO) effective model wherein $c_2=0$ (dotted), and the next-to-leading order (NLO) effective model (dashed). }
  \label{Coulomb_PhaseFactor_1}
  \end{figure}

\begin{figure}[htp]
  \begin{center}
    \includegraphics[scale=.7]{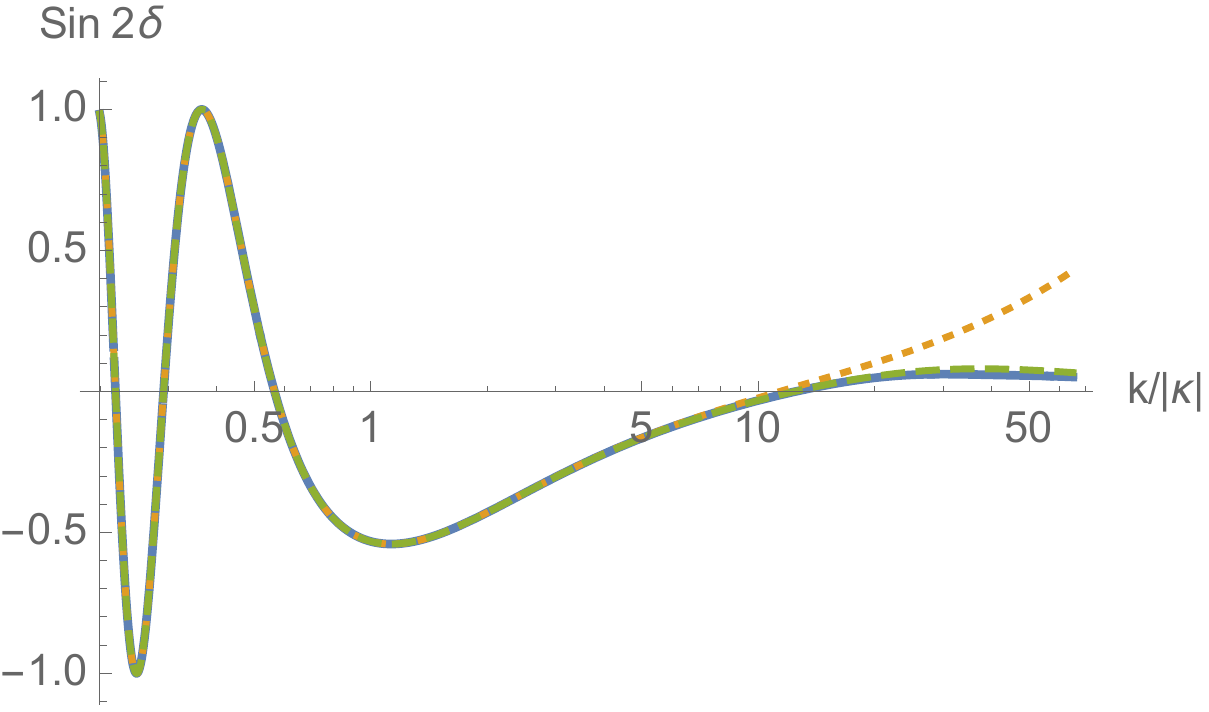}
  \end{center}
  \caption{As in Figure \ref{Coulomb_PhaseFactor_1}, but  $L=0.11 \abs{\k}^{-1}$.}
  \label{Coulomb_PhaseFactor_2}
\end{figure}

The robustness of this method can be tested by predicting the scattering phase shifts and comparing to the predictions from the same UV-complete model.  Using the effective parameters in \eqref{Coulomb_effective_parameters}, inserted into \eqref{Coulomb_EQM_scatt}, the scattering results are obtained; for selected parameters they are summarized in Figures \ref{Coulomb_PhaseFactor_1} and \ref{Coulomb_PhaseFactor_2}. For illustration, the lowest-order (LO) model (equivalent to the self-adjoint extension analysis, in which $c_2=0$) is shown with the next-to-lowest order (NLO) model, which uses the parameters as given in equation \eqref{Coulomb_effective_parameters}.


Given the remarkable agreement between the UV-complete and the effective theory presented here, we note that a similar level of agreement may be achieved in an effective theory that does not exclude the region near $x=0$, but does incorporate a series of momentum-dependent contact potentials in the Hamiltonian. This is done in reference \cite{Lepage:1997cs} for the modified 3-dimensional Coulomb system, wherein results similar to those presented in Figure \ref{RelativeEnergyDiff} may be found. It would seem, therefore, that the ansatz in equation \eqref{Coulomb_effective_ansatz} plays the role of those momentum-dependent contact potentials.


\sec{The $1/x^2$ potential}\label{Sec:1OVERx2}
Consider the system described at long distances by the Hamiltonian
\begin{equation}
H=-\f{\hbar^2}{2m}\f{\d^2}{\d x^2}-\f{a}{x^2}\,.
\end{equation}
Using notation consistent with reference \cite{EssinGriffiths2006}, we set $\hbar=1$ and define
\begin{equation}
\a\equiv2ma\,,
\end{equation}
making note that $\a$ defined here is not the fine-structure constant, as was the case in Section \ref{Sec:1OVERx}. Let
\begin{equation}
2mE\equiv
\begin{cases}
-q^2, &(E<0)\\
k^2, &(E>0)\,,
\end{cases}
\end{equation}
then
\begin{equation}
\(\d_x^2+\f{\a}{x^2}\)\psi=
\begin{cases}
q^2 \psi, &(E<0)\\
 -k^2 \psi, &(E>0)\,.
\end{cases}
\end{equation}
As is well-known, this system has no intrinsic length scale; some non-trivial analysis is needed to compute the bound-state spectrum, as explained in \cite{EssinGriffiths2006}.

For $E<0$, one set of linearly independent solutions is $\sqrt{x} I_{ig}(q x) $ and $ \sqrt{x} K_{ig}(q x)$, where $g\equiv\sqrt{\a-1/4}$ is assumeed to be real; below we will show that $g$ must be real for a bound state to exist, and therefore $\a\geq1/4$ is required. Normalizeability requires that the Bessel-$I$ function be omitted, therefore
\begin{equation}\label{oneOVERx2_sol}
\psi=A \sqrt{x} K_{ig}(q x)\,,
\end{equation}
where $A$ is a normalization factor.  The spectrum of $q$ are observable.

For $E>0$, the Hankel functions are used:
\begin{equation}\label{1overx2_scat}
\psi=A\(\sqrt{x} H^{(2)}_{ig}(k x)-i e^{-g\pi} e^{2i\de}\sqrt{x} H^{(1)}_{ig}(k x) \) \,,
\end{equation}
where the coefficients are chosen such that
\begin{equation}
\lim_{x\to \infty}\psi\sim \f{2\p}{k}e^{-g\pi/2} B\(e^{-ikx}-e^{2i\de}e^{+ikx}\)\,,
\end{equation}
where $2\de$ is the total phase shift.

\ssec{Effective Model}
\sssec{Bound state ($E<0$) solutions}

Consider the application of the boundary condition, equation \eqref{1d_standard_bc}. A series expansion in $q x_\text{b}\ll1$ gives
\begin{equation}
\f{\sqrt{x_\text{b}}}{2}\(\(2+\tilde{Z}+2ig\tilde{Z}\)\(\f{q x_\text{b}}{2}\)^{ig}\Gamma(-ig)+\text{c.c.}\)=0\,,
\end{equation}
where $\tilde{Z}=Z/x_\text{b}$, and is a real function. The complex term in parentheses,
\begin{equation*}
2+\tilde{Z}+2ig\tilde{Z}\,, 
\end{equation*}
has a complex argument
\begin{equation}
\arctan{\f{2g\tilde{Z}}{2+\tilde{Z}}}\,,
\end{equation}
up to some integer multiple of $\pi$. 
Apparently this requires
\begin{equation}\label{1overx2quantization}
g\ln{\f{q x_\text{b}}{2}} + \arctan{\f{2g\tilde{Z}}{2+\tilde{Z}}} + \arg{\Gamma(-ig)}=\(n+\f{1}{2}\) \p\,,
\end{equation}
for an integer $n$. 
The $x_b$-independence of $q$ can be enforced through differentiation of the above with respect to $x_\text{b}$, yielding the differential equation
\begin{equation}
 \tilde{Z}'(x_\text{b})=-\f{1}{4x_\text{b}}\(\(2+\tilde{Z}(x_\text{b})\)^2+\(2g\tilde{Z}(x_\text{b})\)^2\)\,,
\end{equation}
whose solution is
\begin{equation}\label{Zt solution}
\tilde{Z}(x_\text{b})=\f{2}{1+4g^2}\(2g\tan{\[g \ln{\f{b}{x_\text{b}}}\]}-1\)\,.
\end{equation}
Here,  $b$ is a dimensionful constant of integration; however, $b$ is expected to be a function of $q^2$, a point we return to below.

To solve for $q$ one may define
\begin{equation*}
\f{2g\tilde{Z}}{2+\tilde{Z}}  \equiv iw
\end{equation*}
and use the identity
\begin{align}
\arctan{iw} = \f{i}{2}\ln{\f{w+1}{1-w}}\,,  \notag
\end{align}
from which it follows that \eqref{1overx2quantization} may be written
\begin{equation}
g\ln{\f{q x_\text{b}}{2}} + \f{i}{2}\ln{\[\f{2g+i}{2g-i}\(\f{b}{x_\text{b}}\)^{-2ig}\]} + \arg{\Gamma(-ig)}=\(n+\f{1}{2}\) \p\,.\notag
\end{equation}
After simplifying, one may solve for the $n$'th value of $q$:
\begin{equation}\label{kn_eqn}
q_n=\f{2}{b}\exp{ \[\f{1}{g}\(  \(n+\f{1}{2}\) \p 
 + \arctan{\f{1}{2g}}  
  - \arg{\Gamma(-ig)}\)\]  }\,,
\end{equation}
which is $x_\text{b}$-independent and requires real $g$, as advertised. It depends explicitly on $n$ and the integration constant $b$, which can only be determined experimentally or by matching with a UV-complete theory. Consistent with the findings of reference \cite{EssinGriffiths2006}, the ratio of adjacent bound state values of $q_n$ is given by $e^{\pi/g}$. This equation holds for all $q_n x_\text{b}\ll1$, so that its derivation remains valid. That is, equation \eqref{kn_eqn} can be trusted for $n\leq n_\text{max}$ determined by the scale at which the potential deviates from its pure $x^{-2}$ form.

Consider now that in \eqref{Zt solution} the $q$-dependence of the integration function is incorporated by the parameterization
\begin{equation}\label{Zt solution_new}
\tilde{Z}(x_\text{b})=\f{2}{1+4g^2}\(2g\tan{\[g\( \ln{\f{b_0}{x_\text{b}}} + \chi\(q^2\)\)\]}-1\)\,,
\end{equation}
where $b_0$ is a $q$-independent constant. It follows that equation \eqref{kn_eqn} is modified to
\begin{align}\label{kn_eqn_new}
q_n&=q_n^{(0)}\exp{\[- \chi\(q^2\)\]}\,,
\end{align}
where
\begin{equation}\label{EQM_1OVERx2LO}
q_n^{(0)}\equiv\f{2}{b_0}\exp{
 \[
\f{1}{g}
\(
 \(n+\f{1}{2}\) \p 
 + \arctan{\f{1}{2g}}  
  - \arg{\Gamma(-ig)}
\)
\]  
}\,.
\end{equation}

For the class of systems in which an analytic low-momentum expansion is appropriate, one may posit the Taylor series form
\begin{equation}
\chi(q^2)=c_0 + c_2 q^2 +{\cal O}(q^4)\,,
\end{equation}
Note that equation \eqref{Zt solution_new} indicates that one can set $c_0=0$ by appropriate redefinition of $b_0$. For $c_2 q^2\ll1$ we find
\begin{equation}\label{EQM_1OVERx2LO_2}
q_n^{}\simeq q_n^{(0)}\(1-c_2 \(q_n^{(0)}\)^2\)\,,
\end{equation}
which has the $n$-dependent form
\begin{equation}\label{n-dep_structure}
\tilde{a}\, e^{n/g}\(1-\tilde{b}\, e^{2n/g}\)\,,
\end{equation}
for two constants $\tilde{a}$ and $\tilde{b}$. We will compare to this the results of a particular UV-complete model described below.

\sssec{Scattering ($E>0$) solutions}

The function $Z(x_\text{b})$ as derived in the previous section may be used here, with which the boundary function \eqref{Zt solution} gives
\begin{equation}\label{EQM_1overx2_LO_phasefactor}
e^{2i\de}\!=\!i
\f{
\(\f{kb}{2}\)^{ig}\!\(2g-i\)\Gamma\(-ig\)+ e^{\p g}\(\f{kb}{2}\)^{-ig}\!\(2g+i\)\Gamma\(ig\)
}
{
e^{\p g}\(\f{kb}{2}\)^{ig}\!\(2g-i\)\Gamma\(-ig\)+ \(\f{kb}{2}\)^{-ig}\!\(2g+i\)\Gamma\(ig\)
}\,,
\end{equation}
which is arrived at after considerable simplification. This lowest order result may be used to obtain the next-to-leading-order (NLO) result by replacing $b\to b_0\(1 - c_2 k^2\)$, assuming the function $\chi(q^2)$ continues analytically from the bound states to the scattering states, i.e. $\chi(q^2)\to \chi(-k^2)$.


\ssec{A UV-complete model}

Consider a model in which the singular potential is made finite at the origin with a potential cap, parameterized by
\begin{equation}
V(x)= 
\begin{cases}
-\f{a}{L^2},~~~~~&(0\leq x \leq L)\\
-\f{a}{x^2},~~~~~~~&(x>L)\,.
\end{cases}
\end{equation}

For bound states ($E<0$) define 
\begin{align}
q^2&\equiv -2mE\notag\\
\a&\equiv 2ma\notag\\
p^2&\equiv\f{\a}{L^2} -q^2\,,
\end{align}
so that for $x> L$ the solutions are just as in equation \eqref{oneOVERx2_sol},
\begin{equation}
\psi_\text{out}=A \sqrt{x} K_{ig}\(q x\)\,,
\end{equation}
and within $x\leq L$
\begin{equation}
\psi_\text{in}=B \sin{px}\,.
\end{equation}
Matching the wave function and its derivative at $x=L$ can be described within a single matching equation
\begin{equation}\label{1OVERx2FullScattMatch}
\psi_\text{out}'\(L\)\psi_\text{in}\(L\)-\psi_\text{out}\(L\)\psi_\text{in}'\(L\)=0\,,
\end{equation}
which, upon expanding to ${\cal O}\(q L\)^2$ is of the form
\begin{align}
\(\f{q L}{2}\)^{ig}\Gamma\(-i g\)&\[A+B \(q L\)^2\right.\notag\\
&\left.+i\(C+D\(q L\)^2\)\] + \text{c.c.}=0\,,
\end{align}
where the constants
\begin{align}
A&=4\(1+g^2\)\sqrt{\a}\(\sin{\sqrt{\a}}-2\sqrt{\a}\cos{\sqrt{\a}}\)\notag\\
B&=2\(1+g^2-\a\)\cos{\sqrt{\a}}+\(1-2g^2\)\sqrt{\a}\sin{\sqrt{\a}}\notag\\
C&=8g\(1+g^2\)\sqrt{\a}\sin{\sqrt{\a}}\notag\\
D&=-g\[\(4+4g^2-2\a\)\cos{\sqrt{\a}}+3\sqrt{\a}\sin{\sqrt{\a}}\]\,.
\end{align}
This apparently requires
\begin{equation}
g \ln{\f{q L}{2}}+\arg{\Gamma\(-ig\)}+\arctan{\f{C+D\(q L\)^2}{A+B \(q L\)^2}}=\(n+\f{1}{2}\)\pi\,.
\end{equation}
This transcendental equation may be solved perturbatively for small $qL$:
\begin{equation}\label{ABC_1}
q_n^{}\simeq q_n^{(0)}\(1-f(\a) \(q_n^{(0)}L\)^2\)
\end{equation}
where
\begin{equation}\label{ABC_2}
q_n^{(0)}=\f{2}{L}\exp{
 \[
\f{1}{g}
\(
 \(n+\f{1}{2}\) \p 
 - \arctan{\f{C}{A}   }  
  - \arg{\Gamma(-ig)}
\)
\]  
}\,,
\end{equation}
and
\begin{align}\label{ABC_3}
f(\a)&=\f{AD-BC}{A^2+C^2}\,.
\end{align}
Note that the $n$-dependent structure is the same as described in the effective model, equation \eqref{n-dep_structure}.

For scattering states ($E>0$), define 
\begin{align}
k^2&\equiv 2mE\notag\\
\a&\equiv 2ma\notag\\
p^2&\equiv\f{\a}{L^2} +k^2\,,
\end{align}
so that for $x> L$ the solutions are just as \eqref{1overx2_scat}
\begin{equation}
\psi_\text{out}=A\(\sqrt{x} H^{(2)}_{ig}(k x)-i e^{-g\pi} e^{2i\de}\sqrt{x} H^{(1)}_{ig}(k x) \) \,,
\end{equation}
and within $x\leq L$
\begin{equation}
\psi_\text{in}=B \sin{px}\,.
\end{equation}
Matching the wave function and its derivative at $x=L$ is performed using \eqref{1OVERx2FullScattMatch}, from which the phase factor may be solved.

\ssec{Matching the UV-complete \& effective models}

By matching the above UV-complete results, equations \eqref{ABC_1},\eqref{ABC_2}, and \eqref{ABC_3} with that of the effective model, equations \eqref{EQM_1OVERx2LO} and \eqref{EQM_1OVERx2LO_2},  we learn that the effective parameters $b_0$ and $c_2$ are 
\begin{align}\label{1OVERx2effectiveparametermatch}
b_0&=L\, \exp{\[\f{1}{g}\(\arctan{\f{1}{2g}}  +  \arctan{\f{C}{A}}\)\]}\notag\\
c_2&=f(\a)L^2\,.
\end{align}

%

The robustness of the method can be checked, as in the previous section, by predicting the scattering phase shift and comparing it to the result from the same UV-complete model. In Figure \ref{1OVERx2IMPhaseFactor}, $\sin{2\de}$ is plotted as a function of $k$ for the UV-complete model using \eqref{1OVERx2FullScattMatch}, the lowest order (LO) effective model using equation \eqref{EQM_1overx2_LO_phasefactor} with $b\to b_0$, and the next-to-leading order (NLO) model using \eqref{EQM_1overx2_LO_phasefactor} with the replacement $b\to b_0\(1 - c_2 k^2\)$. 

\begin{figure}[htp]
  \begin{center}
    \includegraphics[scale=.77]{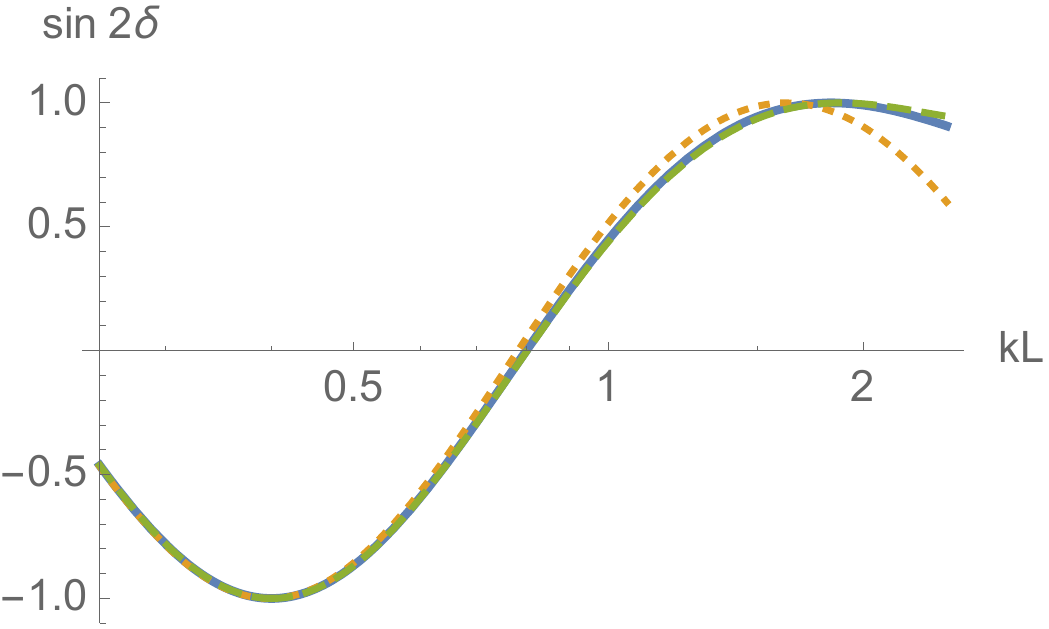}\caption{Scattering results for the $1/x^2$ potential. As a function of the dimensionless product $kL$, $\sin{2\de}$ is shown for $\a=1.5$ as computed in the UV-complete model (solid curve), the lowest order (LO) effective model wherein $c_2=0$ (dotted), and the next-to-leading order (NLO) effective model (dashed).}
  \end{center}
\end{figure}\label{1OVERx2IMPhaseFactor}


\sec{Free particle}\label{sec:free}

The free particle on the real axis is described by the Hamiltonian
%
\begin{equation}
H=-\f{\hbar^2}{2m}\f{\d^2}{\d x^2}\,.
\end{equation}
Let $\hbar=1$ and define
\begin{equation}
2mE\equiv
\begin{cases}
-q^2, &(E<0)\\
k^2, &(E>0)\,.
\end{cases}
\end{equation}
For scattering states
\begin{equation}\label{free_1d_scattering}
\psi =A\sin{(kx +\de)}\,,
\end{equation}
and for bound states
\begin{equation}\label{free_1d_boundstate}
\psi=Ae^{-q x}\,.
\end{equation}

\ssec{Effective Model}
The wavefunction must satisfy the boundary condition in equation \eqref{1d_standard_bc}. For scattering states, the series expansion in small $kx_\text{b}$ indicates
\begin{equation}\label{1d_EQM_tand}
\tan{\de}=-k Z\,.
\end{equation}
%

As in the previous sections, we would generally consider that $Z$ could vary with $x_\text{b}$, but by inspection it clearly does not in this system. On the other hand, $\tan{\de}$ could have a complicated dependence on $k$, indicating that $Z$ may be a function of $k$. Consistent with the Sections \ref{Sec:1OVERx} and \ref{Sec:1OVERx2} above we therefore choose the notation
\begin{equation}\label{Free_Z_to_chi}
Z\to \chi(k^2)\,.
\end{equation}
A perturbative \emph{ansatz} will be made for $\chi$; here, as in other sections, we could posit a simple Taylor series which has proved effective thus far. However; the limitation of the Taylor series becomes apparent when trying to effectively capture resonances in this model, a point that will be addressed below.

For bound states, equation \eqref{1d_standard_bc} gives
\begin{equation}
q=\f{1}{Z}\to\f{1}{\chi(-q^2)}\,,
\end{equation}
having assumed that $\chi(k^2)$ can be analytically continued to negative arguments.


\ssec{A UV-complete model}
Consider a model in which the potential contains a well of small width, $L$:
\begin{equation}
V(x)= 
\begin{cases}
-V_0,~~~~~&(0\leq x \leq L)\\
0,~~~~~~~&(x>L)\,,
\end{cases}
\end{equation}
where $V_0>0$.

For scattering states write
\begin{equation}
E=\f{k^2}{2m}
\end{equation}
and
\begin{equation}
p=\sqrt{k^2+2mV_0}
\end{equation}
so that the spatial part of the exterior solution ($x> L$) is the same as equation \eqref{free_1d_scattering},
\begin{equation}\label{free-UV_scatt_sol_out}
\psi_\text{out}=A \sin{(kx +\de)}\,,
\end{equation}
and within $x\leq L$
\begin{equation}\label{free-UV_scatt_sol_in}
\psi_\text{in}=B \sin{px}\,.
\end{equation}

By matching the wavefunction and its derivative inside and outside the step, one may show that
\begin{equation}\label{1d_UV_tand}
\tan{\de}=\f{-p\tan{kL}+k\tanh{pL}}{p+k\tan{kL}\tanh{pL}}\,.
\end{equation}

For bound states write
\begin{equation}
E=-\f{q^2}{2m}
\end{equation}
and
\begin{equation}
p=\sqrt{-q^2+2mV_0}
\end{equation}
so that the spatial part of the exterior solution ($x> L$) is the same as equation \eqref{free_1d_boundstate}
\begin{equation}
\psi_\text{out}=A e^{-qx}\,,
\end{equation}
and within $x\leq L$ it as in \eqref{free-UV_scatt_sol_in},
\begin{equation}\label{free-UV_bound_sol_in}
\psi_\text{in}=B \sin{px}\,.
\end{equation}

By matching the wavefunction and its derivative inside and outside the step, one may show that in this full model
\begin{equation}\label{1d_UV_q}
q=-\f{p}{\cot{pL}}\,.
\end{equation}

\ssec{Matching the UV-complete \& effective models}

In the long wavelength limit, i.e. for $k$ small relative to  $L^{-1}$ and $\sqrt{2mV_0}$, equation \eqref{1d_UV_tand} may be written as a Taylor series expansion in odd powers of $k$; however, here it is advantageous to use a Pad\'e approximant (see, e.g., \cite{bender1999advanced}), which may be used to perturbatively describe divergent functions. Up to order $k^2$,
\begin{equation}\label{1d_UV_tand_PadeApprox}
\tan{\de}=-k\f{a_0}{1+b_2 k^2}\,,
\end{equation}
where
\begin{equation}
a_0=\(L-\f{\tan{L\sqrt{2mV_0}}}{\sqrt{2mV_0}}\)
\end{equation}
and
\begin{widetext}
\begin{equation}
b_2=\frac{12 L^2 m V_0+L \sqrt{2 m V_0} \tan \left(L \sqrt{2m V_0}\right) \left(\left(3-4 L^2 m V_0\right) \cot
   ^2\left( L \sqrt{2m V_0}\right)-3\right)-3}{12 m V_0 \left(L \sqrt{2m V_0} \cot \left(L \sqrt{2m
   V_0}\right)-1\right)}\,.
\end{equation}
\end{widetext}
Matching the effective model  \eqref{1d_EQM_tand} with \eqref{1d_UV_tand_PadeApprox} requires
\begin{equation}\label{free_chi}
\chi(k^2)=\f{a_0}{1+b_2 k^2}\,.
\end{equation}
%

The bound state(s) as predicted by the effective model should be consistent with equation \eqref{free_chi} for states satisfying $qL\ll1$:
\begin{equation}
q=\f{1}{\chi(-q^2)}=\f{1-b_2 q^2}{a_0}
\end{equation}

\begin{figure}[ht]
  \begin{center}
    \includegraphics[scale=.7]{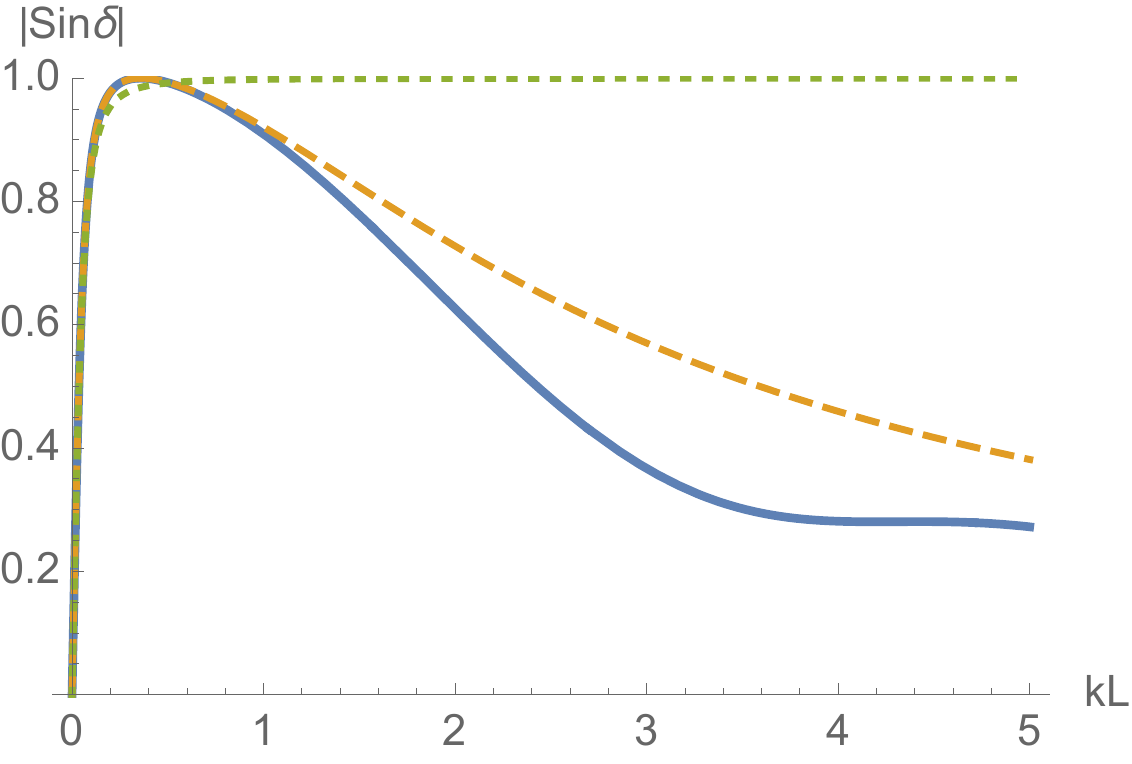}
  \end{center}
  \caption{Scattering results for the free particle for the chosen parameter: $V_0=1.3/(mL^2)$.}
  \label{FreeScatter_1pt3}
\end{figure}
\begin{figure}[ht]
  \begin{center}
    \includegraphics[scale=.7]{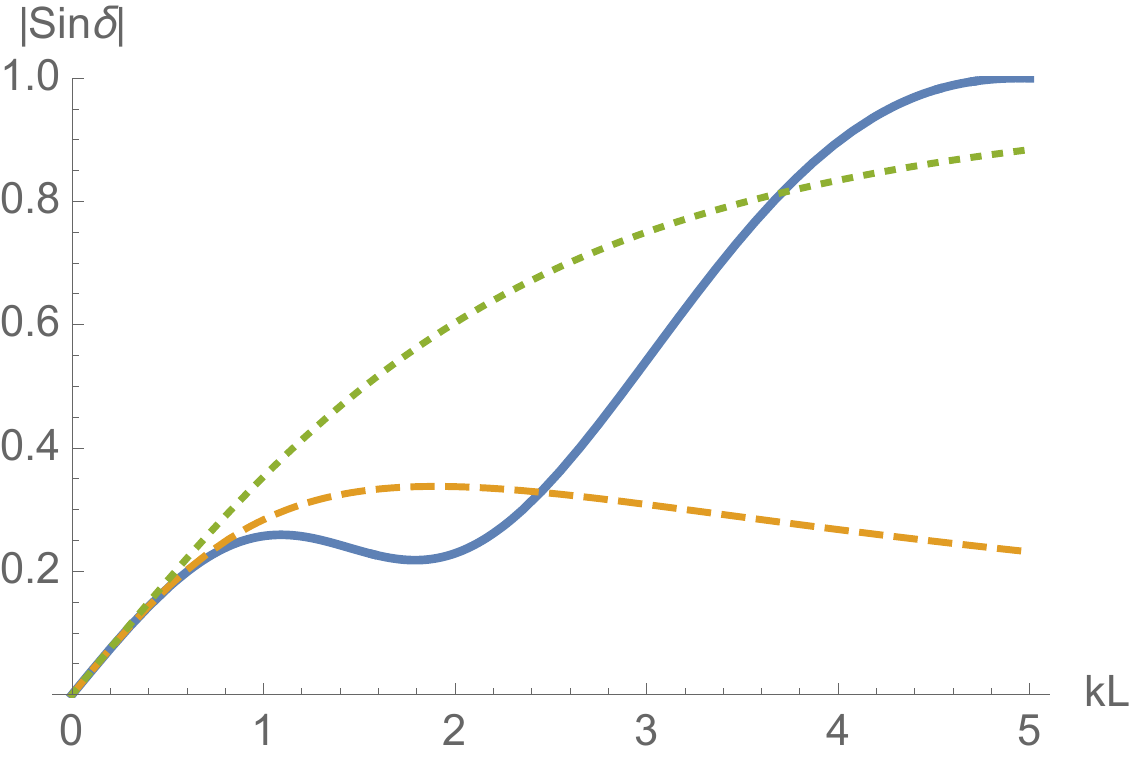}
  \end{center}
  \caption{As in Figure \ref{FreeScatter_1pt3}. Chosen parameter: $V_0=9.5/(mL^2)$.}
  \label{FreeScatter_9pt5}
\end{figure}
\begin{figure}[ht]
  \begin{center}
    \includegraphics[scale=.7]{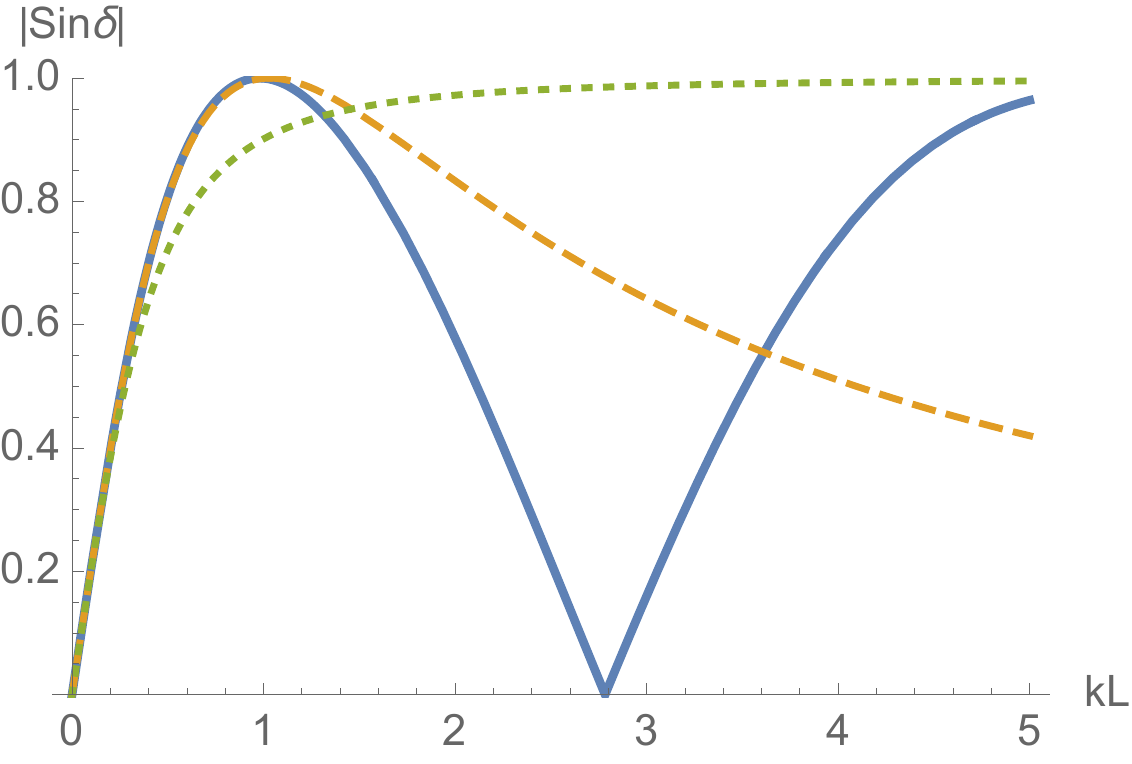}
  \end{center}
  \caption{As in Figure \ref{FreeScatter_1pt3}. Chosen parameter: $V_0=12.0/(mL^2)$.}
  \label{FreeScatter_12pt0}
\end{figure}

In order to show the goodness (or lack thereof) of the effective model, we choose $m=1.0$ and $L=1.0$ (in the appropriate units) and vary $V_0$. In Figures \ref{FreeScatter_1pt3}, \ref{FreeScatter_9pt5}, and \ref{FreeScatter_12pt0} the absolute value of $\sin{\delta}$ is shown as a function of wave number, $k$, for the UV-complete model, the next-to-leading order (NLO) effective effective model, and the lowest order (LO) effective model with $b_2=0$. In Table \ref{1d-free_matching_compare_bound} the results for the least-bound state $q$ is displayed. What is clear from these results is that for very low values of $k$, the effective method is accurate. At higher values of $k$, near the first resonance, the model is only accurate for a range of system parameters such that the resonance occurs at a momentum $\lesssim L^{-1}$.

\begin{table}[ht]
\caption{Values of bound-state $q$ in the free-particle system. Chosen parameters are $m=1.0, L=1.0$, and various values for $V_0$, in units of $(mL^2)^{-1}$. No exact solution can be obtained for $V_0=2.9$; however, the local extremum of the function $q \chi(-q^2)$  gives a best approximation.} 
\centering 
\begin{tabular}{c c c c} 
\hline\hline 
$V_0$ & UV-complete & Effective Model & Fractional Error \\ [0.5ex] 
\hline 
1.3 & 0.0650338 & 0.0650344 & $9.2\times10^{-6}$ \\ 
1.7 & 0.413468 & 0.414602 & $2.7\times10^{-3}$ \\
2.1 & 0.707634 & 0.719503 & $1.7\times10^{-2}$ \\
2.5 & 0.965104 & 1.02214 & $5.9\times10^{-2}$ \\ 
2.9 & 1.19587 & 1.4441 & $0.21$ \\ 
9.5 & 3.55377 & -- & -- \\
12.0 & 0.67975 & 0.704712 & $3.7\times10^{-2}$ \\ [1ex] 
\hline 
\end{tabular} 
\label{1d-free_matching_compare_bound}
\end{table}

\sec{Scattering Time Delays}\label{Sec:WignerDelay}

Finally, we compute the time delay associated with the scattering of a wave packet, following an analysis similar to reference \cite{PhysRevA.66.052102}. The state consisting of an incoming wave packet scattering into an outgoing wave packet may be written as the superposition
\begin{equation}\label{gen_wavepacket}
\Psi\(t,x\)=\int_0^\infty dk \,f(k) \,e^{-i\o t}\[\psi_L - e^{2i\de}\psi_R   \]\,.
\end{equation}


For the free-particle and $1/x^2$-potential systems, equation \eqref{gen_wavepacket} takes the asymptotic form
\begin{equation}
\Psi\(t,x\)\sim\int_0^\infty dk \,f(k) \,e^{-i\o t +i k x_0 }\[e^{-ikx} - e^{+ikx+2i\de}    \]\,,
\end{equation}
as $x\to \infty$, taking the packet to be peaked in real space at $x=x_0$ when $t=0$. Assuming that in momentum space it is peaked at $k=k_0$, the stationary phase approximation indicates that the position of the peak of the outgoing wave packet is
\begin{equation}
x_\text{out}(t)=-x_0 + v_0 t - 2\de'(k_0)\,,
\end{equation}
where the group velocity is
\begin{align}
v_0&\equiv \f{d\o}{dk}\Big|_{k=k_0}\,.
\end{align}
The total of time-of-flight for the reflected (outgoing) wave pack to return to position $x=x_0$ is apparently
\begin{equation}\label{long-distance_free_TOF}
t=\f{2x_0}{v_0} + \t\,,
\end{equation}
where the first term is the classical time-of-flight in the absence of any potential; the second term results from the wave packet interaction with the potential, including whatever short-distance interactions occur in the vicinity of $x=0$, and may be written
\begin{align}\label{time_delay_eqn}
\t&=\f{2\de'(k)}{v_0}\notag\\
&=2\f{\d\de(E)}{\d E}\,.
\end{align}
This is referred to as the Wigner time delay \cite{Wigner:1955zz}.

An analytic description of the delay near a sharp resonance is illuminating for the case of the free particle. From equation \eqref{1d_UV_tand_PadeApprox} it may be shown that
\begin{align}
e^{2i\de}&=\f{1+b_2k^2-ika_0}{1+b_2k^2+ika_0}  \notag\\
&=\f{E-E_r+i\sqrt{2mE}a_0 E_r}{E-E_r-i\sqrt{2mE}a_0 E_r}\,,
\end{align}
where $E_r=-1/(2mb_2)$. In the vicinity of a sharp resonance, under the condition $\abs{a_0}/\sqrt{\abs{b_2}}\ll1$, we have
\begin{equation}
e^{2i\de}\simeq\f{E-E_r+i\f{\Gamma}{2}}{E-E_r-i\f{\Gamma}{2}}\,,
\end{equation}
where $\Gamma=\sqrt{8m}a_0 E_r^{3/2}$. It follows that
\begin{equation}
\t\simeq -\f{\Gamma}{\(E-E_r\)^2+\(\f{\Gamma}{2}\)^2}\,.
\end{equation}
This analysis is not possible with a (lowest-order) self-adjoint extension, as considered in reference \cite{PhysRevA.66.052102}, wherein only pure contact potentials were considered, i.e. $b_2=0$.

 
For the one-dimensional Coulomb system, equation \eqref{gen_wavepacket} takes the asymptotic form
\begin{align}
\Psi\(t,x\)\sim\int_0^\infty dk &\,f(k) \,e^{-i\o t +i k x_0 -i\f{\k}{k}\ln{2kx_0}} \notag\\
&\times  \[e^{-ikx + i\f{\k}{k}\ln{2kx}} - e^{2i\de}e^{+ikx - i\f{\k}{k}\ln{2kx}}      \]\,,
\end{align}
as $x\to \infty$, again taking the incoming wave packet to be peaked in real space at $x=x_0$ when $t=0$. Here, the stationary phase approximation indicates that the peak of the reflected wave packet arrives at $x=x_0$ at the time
\begin{equation}\label{long-distance_Coulomb_TOF}
t=\f{2}{v_0}\(x_0 -\f{\k}{k^2}+\f{\k \ln{2kx_0}}{k^2} \) + \t\,,
\end{equation}
where $\t$ is as defined in equation \eqref{time_delay_eqn}. The additional terms in equation \eqref{long-distance_Coulomb_TOF}, compared to equation \eqref{long-distance_free_TOF}, are due to the long-range nature of the Coulomb potential.

\sec{Instantaneous vs. Time-averaged Quantities}\label{Sec:Consequences}

The above sections have demonstrated the utility of the proposed effective method. The consequences of this proposal is that Hamiltonian fails to be hermitian, states fail to be orthogonal, and probability is not conserved for infinitesimal translations in time; however, all the canonical relations hold in a time-averaged sense. Thus the terms \emph{instantaneous} and \emph{time-averaged} will distinguish between the two cases.

Here we use the generic Hamiltonian specified in equation \eqref{generic_Ham}, to which the eigenfunctions of the Schrodinger equation are of the form
\begin{equation}
\Phi_i(x,t)=e^{-iE_i t}\psi_i(x)\,.
\end{equation}
The wavefunctions are presumed to be well-behaved\footnote{For sake of argument, assume that there is a discrete set of modes living a box of size $D$ which is very large; let the boundary conditions be that all $\psi_i(D)=0$.} in the $x\to\infty$ limit; however, a boundary condition is required at $x=x_b$ given by equation \eqref{1d_standard_bc}:
\begin{equation*}
\psi_i(x_\text{b}) + Z_i(x_b)\, \psi_i'(x_\text{b})=0\,,
\end{equation*}
where, canonically, the function $Z_i$ would be \emph{independent} of a particular mode, $i$. This would be sufficient to ensure eigenmodes with distinct eigenvalues are orthogonal, the Hamiltonian is hermitian, and the evolution is unitary. 

If the boundary function depends on momentum, each mode ``feels" a different function $Z_i$. Consider two distinct eigenfunctions $\Phi_i(x,t)$ and $\Phi_j(x,t)$. The inner product between these two such states is
\begin{equation}
\<\Phi_i,\Phi_j\>=\int_{x_b}^\infty dx\, \Phi^\dagger_i\Phi_j\,.
\end{equation}
The quantity
\begin{align}\label{1d_hermitian_check}
\<H\,\Phi_i,\Phi_j\>\! &-\! \<\Phi_i,H\,\Phi_j\>\notag\\
&=-\f{1}{2m}\!\(\Phi_i^\dagger \f{\d\Phi_j}{\d x}\!-\!\f{\d\Phi_i^\dagger}{\d x} \Phi_j\)\!\bigg|_{x=x_b}\notag\\
&=\f{1}{2m}\(Z_i - Z_j\)\f{\d\psi_i^\dagger}{\d x}\f{\d\psi_j}{\d x}e^{i\(E_i-E_j\)t}\bigg|_{x=x_b}\,.
\end{align}
The necessary and sufficient condition for $H$ to be exactly, or \emph{instantaneously} hermitian is for this quantity to vanish, which is not the case unless $Z_i$ is identically equal to $Z_j$. However, one should note two key features: (1) this quantity time-averages to zero over the period $2\p/\(E_i-E_j\)$ and (2) the amplitude of the ``non-hermicity" is controlled by the difference $Z_i-Z_j$ which, for $E_j$ sufficiently close to $E_i$, will scale as  $E_i-E_j$ raised to some power\footnote{This assumes that $Z_i=Z(E_i)$ is an analytic function of $E_i$.}. 

For real eigenvalues $E_i$ and $E_j$, a textbook analysis indicates that from the violation of instantaneous hermiticity, equation \eqref{1d_hermitian_check}, follows a lack of instantaneous orthogonality:
\begin{equation}
\<\Phi_i,\Phi_j\> =  \f{1}{2m}\f{\(Z_i - Z_j\)}{E_i-E_j}\f{\d\psi_i^\dagger}{\d x}\f{\d\psi_j}{\d x}e^{i\(E_i-E_j\)t}\bigg|_{x=x_b}\,,
\end{equation}
which also time-averages to zero over sufficiently long times for all $i\neq j$.

Finally, consider a state $\Upsilon$ that is a linear combination of $\Phi_i$ and $\Phi_j$, written as 
\begin{equation}
\Upsilon=c_i\Phi_i+c_j\Phi_j\,,
\end{equation}
where $c_i$ and $c_j$ are time independent constants. The inner product is therefore
\begin{align}
\<\Upsilon,\Upsilon\>=\abs{c_i}^2 &\<\Phi_i,\Phi_i\>+\abs{c_j}^2\<\Phi_j,\Phi_j\>\notag\\
&+c_j^*c_i\<\Phi_j,\Phi_i\>+ c_i^*c_j\<\Phi_i,\Phi_j\>\,.
\end{align}
Since the time derivative of the inner product between eigenmodes is
\begin{align}
\f{d}{dt}\<\Phi_i,\Phi_j\>&=\int_{x_b}^\infty dx\, \(\f{\d\Phi_i}{\d t}\)^\dagger\Phi_j+\Psi^\dagger_i\f{\d\Phi_j}{\d t}\notag\\
&=\int_{x_b}^\infty dx\, \(-iH\Phi_i\)^\dagger\Phi_j+\Psi^\dagger_i\(-iH\Phi_i\Phi_j\)\notag\\
&=i\(\<H\Phi_i,\Phi_j\>-\<\Phi_i,H\Phi_j\>\)\,,
\end{align}
it follows that
\begin{align}
\f{d}{dt}\<\Upsilon,\Upsilon\>&=c_i^*c_j\f{d}{dt}\<\Phi_i,\Phi_j\> + \text{c.c.}\notag\\
&=\rho_{ij}\(Z_i-Z_j\) \cos{\[\(E_i-E_j\)t+\th_{ij}\]}\,,
\end{align}
where
\begin{equation}
\rho_{ij}\equiv \abs{c_i^*c_j \f{i}{m}\f{\d\psi_i^\dagger}{\d x}\f{\d\psi_j}{\d x}}
\end{equation}
and
\begin{equation}
\th_{ij}\equiv \arg{\rho_{ij}} \,.      
\end{equation}
Therefore, although the time derivative of the norm of this composite state is not zero, it oscillates in time at a frequency of $\(E_i-E_j\)/2\p$, time-averages to zero, and has vanishing amplitude in the limit $E_j\to E_i$. 

Apparently, these canonical quantum mechanical relations, and others that are derived from them, are obeyed if the usual instantaneous inner products are replaced with their time-averaged versions:
\begin{equation}
\<A,B\> \to \<A,B\>_T \,, 
\end{equation}
for generic states $A$ and $B$, where 
\begin{equation}
\<A,B\>_T \equiv \f{1}{T} \int_{-T/2}^{T/2}\,dt\, \<A,B\>\,,
\end{equation}
where $T$ is longer than the minimum required averaging time. There is a class of real systems in which the experimental time resolution is much greater than $T$, in which case unitarity violation is not observable, and the method described herein has predictive power.

\sec{Discussion}\label{Sec:Discussion}


Here we have proposed a method for constructing an effective long-distance quantum mechanical description of systems in which small regions of space are omitted from analysis; in other words, the region of analysis is bounded artificially.   With this method, a free function -- here called $\chi$ -- arises from the requirement that observables do not depend on the location of the artificial boundary. It appears that, at least for a certain class of stationary systems, $\chi$ can be described by an approximant in the variable $q^2$ for bound states ($-k^2$, for scattering states). Therefore, this is a method to perturbatively resolve contact potentials.

The robustness of this effective method has been demonstrated for potentials that have the long-distance scaling of $1/x$, providing a new perspective on the theory of quantum defects in one dimension, and is also applicable for potentials of the form $1/x^2$, and for free particles. Furthermore, the Wigner time delay associated with a sharp resonance can be computed with this method.

In subsequent work, this technique will be applied to higher-dimensional systems of contemporary interest.  Applied to three-dimensional hydrogen-like atoms, it may provide a new perspective with which to view the proton radius puzzle \cite{Pohl:2013yb}. It also appears to be applicable to relativistic systems, including those described by the two-dimensional Dirac equation, such as graphene. This may provide a reliable way to incorporate the short-distance, non-relativistic interactions of electrons with their long-distance, effectively massless description.
\\\\\\\\\\\\\\
\begin{center}
{\bf Acknowledgements}\\
\end{center}
Many thanks are owed to Harsh Mathur, who introduced me to the subject of self-adjoint extensions and provided useful feedback during the early stages of this work. I would also like to thank Gwyneth Allwright, with whom I've had many constructive discussions about the applications of boundary conditions in quantum mechanics. Additionally, I would like to thank Kate Brown for useful discussions. Lastly, I would like to thank the students of BBHHS and CFHS who kept me interested in physics, and for whom it was a great pleasure to teach.

\bibliography{EQM_bib}
\end{document}